\documentclass{jltp}

\usepackage{graphicx} 

\title{Domain State Occurring
in the de Haas-van Alphen Effect in Silver}

\author{J.~L. Smith and J.~C. Lashley}

\address{Los Alamos National Laboratory, Los Alamos, New Mexico 87545, USA\\}

\runninghead{J.~L. Smith and J.~C. Lashley}{Domain State Occurring
in the de Haas-van Alphen Effect in Silver}

\begin{document}

\maketitle

\begin{abstract}
Hysteresis has been observed in de Haas-van Alphen measurements of
the Condon domains in silver, and it shows the first-order
nature of the transition to the domain state. The hysteresis, and
thus the first-order nature, is manifested in a nonlinear effect
where  a double-valued response of the amplitude with the applied
external field is observed.

\end{abstract}

\section{INTRODUCTION}

The de Haas-van Alphen (dHvA) effect is perhaps the most
significant of the quantum oscillatory phenomena in metals; as the
magnetization $\emph{M}$ is a thermodynamic state function, it can
be directly related to the density of states and Fermi-Dirac
distribution of electrons in a metal.  In order to measure this
effect, one observes the oscillatory magnetization with swept field of an
electrically conducting single crystal at a low temperature and
high-magnetic field.  In the course of measuring the Fermi
surfaces of the elements, copper, silver, and gold oscillations
came late because both higher fields and better samples were
required\cite{Shoenberg}.  The large spheres of their Fermi
surfaces barely touch each other in reciprocal space and in turn
give rise to closely spaced belly oscillations.  Although the
periods seemed correct, the oscillations were distorted by
harmonic content.  An explanation was provided by Shoenberg and
Pippard who realized that if the magnetization $M$ was
swinging by more than the period of the oscillation, the internal
field \emph{B} = \emph{H} + 4$\pi\emph{M}$ must be used to
modify the oscillatory equation to obtain \cite{Shoenberg}
\begin{equation}
\emph{M}= A~\sin\left(\frac{2\pi F}{H +4\pi M}  + \phi\right)
\end{equation}
where the variables are as in Shoenberg\cite{Shoenberg}. This
description seemed to be in favorable agreement with experiment,
and then Pippard realized that the effective free energy became
multi-valued at high enough fields\cite{Shoenberg}. Consequently,
this result leads to coexisting phases as in the pressure-volume
isotherms in gas-liquid phase equilibria, and in the case of
silver, it is Condon domains with slightly different
magnetizations. NMR experiments indicated that such
domains differ, for example, by about a mT at 9 T. This
observation led to the name Condon domains\cite{Condon2}.
Motivated by recent review articles on this subject\cite{Solt,Gordon}, we describe the thesis work of one of the
authors (JLS) under the direction of George Seidel\cite{Smith}.

\section{EXPERIMENT}
\begin{figure}
\centering
\includegraphics[height=2.5in]{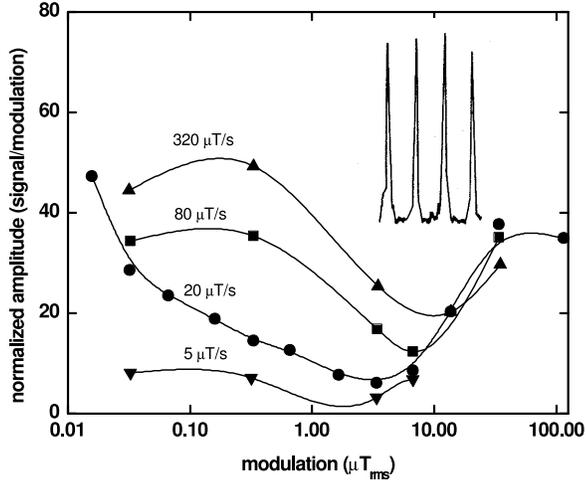}
\caption[]{The signal amplitude divided by the modulation level
 at various modulations and sweep rates of the applied magnetic field at $T=0.15$ K. The
inset, typical trace is at 0.10 K, 80 $\mu$T/s sweep, and 0.33 $\mu$T$_{\textrm{rms}}$ modulation. The modulation frequencies were 24 Hz.}
\end{figure}

The home-built dilution refrigerator was capable of 20 mK in an 8
T superconducting magnet, which had a homogeneity of one part in
10$^5$ over a 25 mm sphere. For this field-modulation measurement
a sample holder was used that rotated in two directions.  The
silver single crystal was cut into a shape approximating an
ellipsoid with principal axes lengths of 11 mm, 4.6 mm, and 0.61
mm, which were the $<$100$>$ directions of the sample. The
measurements described here are only for the orientation of the
long axis 13.5$^\circ$ away from the field direction and in a
$<$100$>$ plane, which yields a single dHvA oscillation from the
belly of the sphere of the Fermi surface. The results for the
$<$100$>$ orientation, which has two orbits, are covered in detail
elsewhere\cite{Smith}. The crystal had a residual resistivity
ratio of 660 and a Dingle temperature of 0.8 K.

\section {RESULTS AND DISCUSSION}
\begin{figure}
\centering
\includegraphics[width=10cm]{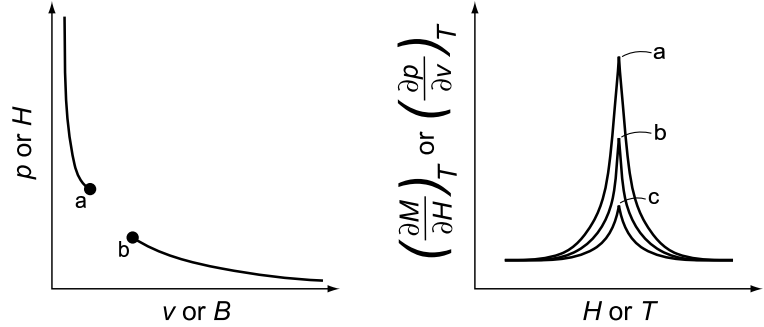}
\caption[]{Pressure-volume or magnetization-field isotherms and
their derivatives are shown for a first-order transition. Because
of magnetoelastic coupling, the work involved in creating and
moving the domain is illustrated as the magnetic analog to
pressure-volume work.} \label{figure2}
\end{figure}

In the situation under investigation we show a rate-dependent
hysteresis effect in Fig. 1. The evolution of the signal amplitude
divided by the modulation level is shown over four decades of
modulation levels. The amplitude for different sweep rates,
ranging from 5 $\mu$T/s to 320 $\mu$T/s, show similar hysteretic
behavior. One can see the existence of the van der Waals
field-induction diagram for Condon domains, a topic covered in
detail by Gordon and coworkers\cite{Gordon}.
If no nonlinearities were present, all of the data would collapse
to one horizontal line. In Ref. \onlinecite{Smith} this was explained as likely due
to domain-wall pinning, which was described as an \emph{ad hoc}
model\cite{Shoenberg}.  Now it is clear that the domain state must
result from a first-order phase transition and that the
nonlinearities arise from the hysteresis of forming and moving the
domains. Because the domains are formed and moved, it is reasonable
to draw parallels between pressure or magnetization versus volume or
field work, respectively, as shown in Fig. \ref{figure2}.
The discontinuous appearance in the dHvA data is similar to
experimental pressure-volume isotherms observed in gas-liquid
phase equilibria. The curves in Fig. 2 illustrate magnetic
work and pressure-volume work. At the discontinuity there is a
similarity to the double-valued response of the amplitude with the
applied external field. The derivatives are plotted in Fig. 2 to
illustrate the rate dependence of the data. In general we observe
the maximum value of the derivative with the lowest modulation level. In many respects this is similar to measuring the height
of the specific-heat peak through a first-order transition\cite{Lashley}. Generally one will take small increments of
temperature to obtain the total height of the transition. We
conclude that the nonlinear dHvA effect can be used as a tool to
unambiguously identify a first-order transition. The multi-valued
amplitude as a function of field is a hallmark of the hysteresis,
and we speculate that the dampening of the amplitude is related to
the latent heat as observed independently in specific-heat measurements\cite{Gordon}.

\section*{ACKNOWLEDGMENTS}

When this work was underway, Mar\'{\i}a Elena de la Cruz,  as a postdoc, was
instrumental to its completion.   Her husband Paco, a
postdoc with Manuel Cardona across the hall, simply could not keep
away from that work.  So Mar\'{\i}a Elena and Paco were crucial to
making the work described here possible and have never been
forgotten.  This work was supported by NSF, DARPA, and the U.S. Dept. of Energy.

\end{document}